# Photodetectors based on junctions of Two-Dimensional Transition metal dichalcogenides

Xia Wei (魏侠), Fa-Guang Yan (闫法光), Chao Shen (申超), Quan-Shan Lv (吕全山) and Kai-You Wang[1,2]† (王开友)

[1]*State Key Laboratory of Superlattices and Microstructures, Institute of Semiconductors, Chinese Academy of Sciences, Beijing 100083, China*

[2]*College of Materials science and Opto-Electronic Technology, University of Chinese Academy of Science*

Transition metal dichalcogenides (TMDCs) have gained considerable attention because of their novel properties and great potential applications. The flakes of TMDCs not only have great light absorptions from visible to near infrared, but also can be stacked together regardless of lattice mismatch like other two-dimensional (2D) materials. Along with the studies on intrinsic properties of TMDCs, the junctions based on TMDCs become more and more important in applications of photodetection. The junctions have shown many exciting possibilities to fully combine the advantages of TMDCs, other 2D materials, conventional and organic semiconductors together. Early studies have greatly enriched the application of TMDCs in photodetection. In this review, we investigate the efforts in photodetectors based on the junctions of TMDCs and analyze the properties of those photodetectors. Homojunctions based on TMDCs can be made by surface chemical doping, elemental doping and electrostatic gating. Heterojunction formed between TMDCs/2D materials, TMDCs/conventional semiconductors and TMDCs/organic semiconductor also deserve more attentions. We also compare the advantages and disadvantages of different junctions, and then give the prospects for the development of junctions based on TMDCs.

## 1. Introduction

Since the discovery of graphene by Geim and Novoselov,[1,2] many unique properties[3-7] have emerged which result in considerable attention on layered materials.[8-11] Many efforts have been put into the fundamental physical researches on properties of layered materials for next generation of electronics and optoelectronics.

* Project supported by '973 Program' No. 2014CB643903, and NSFC Grant No. 61225021, 11474272, 11174272 and 11404324. The project was also sponsored by K. C. Wong Education Foundation.
† Corresponding author. E-mail: kywang@semi.ac.cn

Until now there have already been hundreds of layered materials that can be thinned down to monolayers and retain their stability.[3] Each layered material has its own merits and demerits. For example, the carrier mobility in black phosphorus (BP) is higher than that in Transition metal dichalcogenides (TMDCs), but TMDCs have larger light absorption compared with BP in visible range.

TMDCs have the general formula MX$_2$ where M is a transition metal (such as Mo, W) and X is a chalcogen atom (such as S, Se, Te).[12-15] The atoms forming bounds are arranged into planes (layers). The layers are combined by van der Waals interactions to form bulk crystals.[16] The weak out-of-plane interactions allow the isolation of single layers by micromechanical exfoliation.

Crystals of TMDCs have been investigated for more than five decades.[16,17] However, only in past decade, the thin layers of TMDCs were set off an upsurge, where many interesting properties were observed, such as charge density wave, superconductivity, variation of band structure and so on.[12,18,19] Among them, the variation of band structure is prominent in optical properties. For example, the bandgaps of TMDCs change from indirect to direct when the TMDCs are thinned down to monolayers. The direct bandgap leads to increased probability for photon to generate electron-hole pairs which results in better light absorptions. Moreover the bandgaps of thin layered TMDCs rely on the number of layers,[12,13,20-24] which allows the absorption of light at different wavelength. Due to the van der Waals interactions between layers, we can also stack different types of TMDCs flakes regardless of lattice mismatching to have different wavelength light absorption.[25]

Earlier, photodetectors based on field-effect transistor (FET) structure of TMDCs have been studied[26-31] and have shown excellent performances such as the highest photoresponsivity of 880 A$\cdot W^{-1}$ based on monolayer MoS$_2$.[31] There have already been many reviews.[24,25,32-36] Because the built-in electric field can effectively separate photogenerated electron-hole pairs, the junctions of TMDCs have gained more attentions for optoelectronics in these years. With the rapidly development of the photodetectors based on the junctions of TMDCs, it is time to review the recent achievements. In this review, we mainly present photodetectors based on junctions of TMDCs flakes utilizing photoconductive effect and photovoltaic effect. Firstly the underlying physics of photodetection based on P-N junctions are briefly introduced. We then will discuss and compare the different kinds of junctions based on TMDCs (including techniques and performances). After that, we conclude the advantages and disadvantages of different junctions. Finally we give the prospects for the future developments of junctions based on TMDCs.

## 2. Two main mechanism of photodetection: photoconductive effect and photovoltaic effect

Different materials normally have different work functions and Fermi levels. When junctions are formed between p type and n type materials, the band structures will change due to the alignment of the Fermi levels. The conduction and valance band edge of p type materials will bend down. On the contrary, the conduction and valance band edge of n type materials will bend up. Then the energy barrier will be generated between the two materials.

In the photoconductive effect, when the photon energy is higher than the bandgap, photon absorptions generate extra electron-hole pairs. The photogenerated electrons and holes can be drifted in opposite directions by built-in electric field and external source-drain voltage, which greatly reduces recombination of electron-hole pairs and generates photocurrent (Fig. 1a). The photocurrent adds to the dark current, resulting in the reduction of the resistance. Under negative external voltage, the internal electric field will be enhanced and the energy barrier will increase by $qV_R$ (Fig. 1a). Photogenerated electrons are easier to be drifted to low energy side (the side of n type material) and photogenerated holes are easier to be drifted to low energy side (the side of p type material). Thus the photogenerated electrons and holes are separated more effectively.

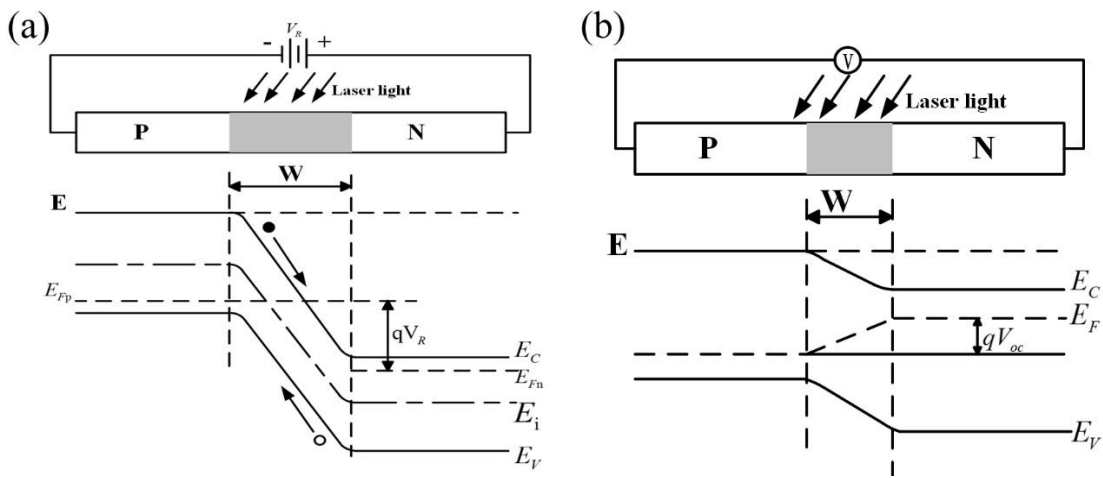

**Fig. 1.** (a) Band alignment of PN junction under photoconductive mode. (b) Band alignment of PN junction under photovoltaic mode.

In the photovoltaic effect, photocarriers are also generated in the same way as the photoconductive effect. But the difference lies in that photogenerated electron-hole pairs are separated only by built-in electric filed rather than external field $V_{ds}$. The built-in electric field generally exists in PN junctions or Schottky junctions. The built-in electric field can effectively separate electrons and holes, which leads to a sizeable short-current $I_{sc}$ at $V_{ds} = 0$. Moreover when the circuit is open, photocarriers will be accumulated on both sides, and open circuit voltage $V_{oc}$ can be obtained (Fig. 1b).

There are two operating modes for the PN junction photodetectors: photoconductive mode (the photodetectors work under external bias, Fig. 1a) and photovoltaic mode (the photodetectors work without external electric field, Fig. 1b). The advantage of photovoltaic mode is the lowest dark current which improves the specific detectivity. The advantages of the photoconductive mode are as follows: First, the reduction of the junction capacitance increases the response speed. Second, under reverse bias, the recombination probability of the photo-generated electrons and holes are smaller, lead it suitable for weak light detection. Last but not least, under reverse bias, the width of depletion layer will be extended, which will increasing photon collection efficiency.

To facilitate comparison of the photodetectors, we mainly focus on the following figures of merits of photodetectors based on junctions: external quantum efficiency (EQE), responsivity (R), specific detectivity (D*) and response time (τ).

External quantum efficiency is a ratio between the number of the charge carriers generated and the number of incident photons on photodetectors. It depends on both the absorption of light and collection of the charges. Responsivity is the ratio between the photocurrent and the incident light power density on photodetectors, which implies the achievable electrical signal under certain illumination. Specific detectivity is equal to the reciprocal of noise-equivalent power, normalized per square root of the sensor's area and frequency bandwidth, which indicates the measure of detector sensitivity. The time response of detectors is measured between 10% and 90% of the final output signal, either on the rising or falling edge.

## 3. Photodetectors based on homojunctions of TMDCs

In conventional semiconductors, homojunctions are the semiconductor interfaces that occur at the similar semiconductor materials with equal bandgaps and different doping. Similarly homojunctions exist at the interface of 2D materials such as graphene,[37-39] WSe₂,[40-43] black phosphorus.[44] In conventional semiconductors, elemental doping is the most common way to form the homojunctions. However, there are three ways to modulate the carrier type of TMDCs to create homojunctions. The first one is to change the surface properties via chemical treatment.[45-48] And the second one is to change the properties of the single-crystalline TMDCs by elemental doping.[49,50] The homojunctions can also be created by electrostatic gating except for doping.[40-43] In the following, we will review the recent development concerning the homojunctions created by these three methods respectively.

## 3.1. Homojunction fabricated by surface chemical doping

In 2012, Hui *et al*. obtained p-doped WSe₂ using physisorbed NO₂. But the doping decays almost within an hour because of the weak physisorption process.[51] Next year they realized n-doping of WSe₂ and MoS₂ by Pstassium which could decrease the contact resistance.[45] Surface chemical doping is demonstrated to lead to the conduction transition of thin TMDC layers. Lately lateral homojunction devices based on surface chemical doping were presented by Choi *et al*. in Fig. 2a.[46] The stacks of MoS₂ and boron nitride (BN) flakes in the devices rely on the mechanical exfoliation and transfer. BN was partially covered on MoS₂ by polydimethysiloxane stamp. The AuCl₃ chemical dopant was spin-coated onto the partially stacked h-BN/MoS₂ heterostructure.

The Au nanoaggregates can form due to the reduction of $AuCl_4^-$ ions in the AuCl₃ solution. Because of the reaction, the ions which are in direct contact with MoS₂ can receive electrons from MoS₂ layers. Thus the p-type MoS₂ will be formed by surface charge transfer, and then the homojunctions of P-N junctions based on MoS₂.

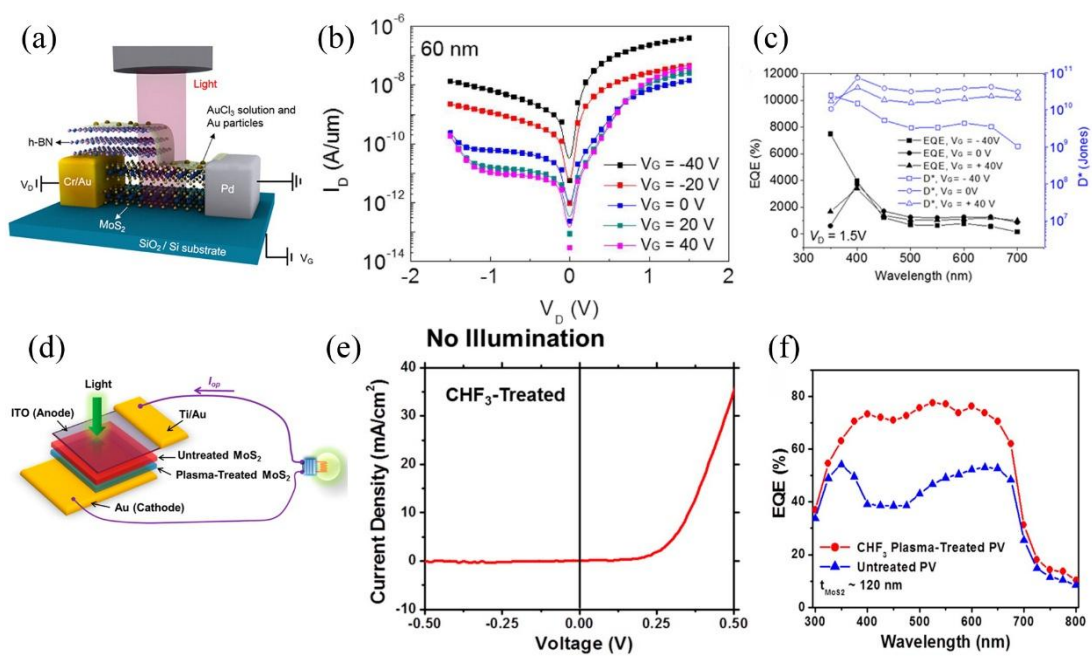

**Fig. 2.** (a) Lateral device schematic diagram of PN junction by AuCl₃ chemical dopant.[46] (b) $I_{ds}$-$V_{ds}$ characteristics measured in different gate configurations.[46] (c) EQE and D* with respect to the wavelength of light at Vds = +1.5 V and Vg = 0, 40 V and -40 V.[46] (d) Vertical PN junction schematic with CHF₃ plasma-treated MoS₂ layers.[52] (e) $J_{ds}$-$V_{ds}$ characteristics of CHF₃ plasma-treated PN junctions.[52] (f) EQE under different wavelength of light.[52]

The log-plotted $I_{ds}$-$V_{ds}$ curves under different gate configurations show the reverse current can be decreased with increasing $V_g$, which is very useful for photodetectors. The large tunable conductance can be attributed to the gating tuning Fermi surface towards or backwards the valance band, which changes the hole transport. Meanwhile the curves in Fig. 2b imply the rectifications resulting from the chemical doping.

Fig. 2c plots the EQE (left axis) and D* (right axis) for different wavelength of light. The maximum EQE obtained is ~7000% at $V_{ds}$ = 1.5 V, $V_g$ = -40 V and 350 nm. It is worth noting that the EQE under different conditions is around 100 times higher than that of the intrinsic MoS₂ phototransistor.[53] Meanwhile the obtained D* is similar to the reported multilayer intrinsic MoS₂ phototransistors. Therefore the AuCl3 doping and stacking processes have not influenced the sensitivity of the devices.

Another example presented here is the MoS₂ P-N junctions fabricated by CHF₃ plasma doping.[52] As shown in Fig. 2e, the junction exhibits more prominent diode-like behavior than the Au/untreated MoS₂/ITO heterostructure. The junction formed at CHF₃ treated and untreated MoS₂ surfaces shows the built-in potential of 0.7eV. Under photovoltaic mode, the EQE of CHF₃ treated MoS₂ PN junction can reach a high value of 78%.

In 2016, Lei *et al.* fabricated planer p-n junctions based on 2D InSe by making use of the long pair electrons found in most of 2D metal chalcogenides rather than relying on lattice defects and physisorption methods.[48] Via a Lewis acid-base reaction, the long pair electrons of n type InSe can react with $Ti^{4+}$ to form p type coordination complexes. Moreover the conduction type conversion in other 2D materials can also be realized via Lewis acids.

To conclude, surface chemical doping is an effective way to prepare photodetectors based on the homojunctions of TMDCs without influencing the sensitivity of the photodetectors. Not only under photoconductive mode but also under photovoltaic mode, the homojunctions have shown higher EQE than the intrinsic TMDCs. Moreover chemical surface doping can also be used to decrease the contact resistance between TMDCs and metal. However, the properties of the homojunctions of TMDCs made by surface chemical doping can easily be affected by external environment. Thus the stability of these devices is rather poor.

### 3.2. Homojunctions fabricated by elemental doping

To have more stable and controllable PN junction, elemental doping has been used to create the homojunctions. Practical applications require substitution of host atoms with dopants, where the doping is secured and stabilized by covalent bonding inside the lattice. Suh *et al.*[49] realized stable P-type doping in bulk $MoS_2$ by substitutional niobium (Nb) and achieved degenerate hole density of $\sim 3 \times 10^{19} cm^{-3}$. In order to form p-type $MoSe_2$, Jin *et al.* also used Nb with five valance electrons to replace Mo which has six valance electrons.[50] Via Nb doping, n-type $MoSe_2$ single crystal can be successfully converted to p-type (Fig. 3a). The thin $n\text{-}MoSe_2$ and $p\text{-}MoSe_2$ layers were stacked by mechanical exfoliation and transfer. Then the vertical homojunctions based on layered $MoSe_2$ have been realized on $SiO_2/Si$ substrates (see insets in Fig. 3b).

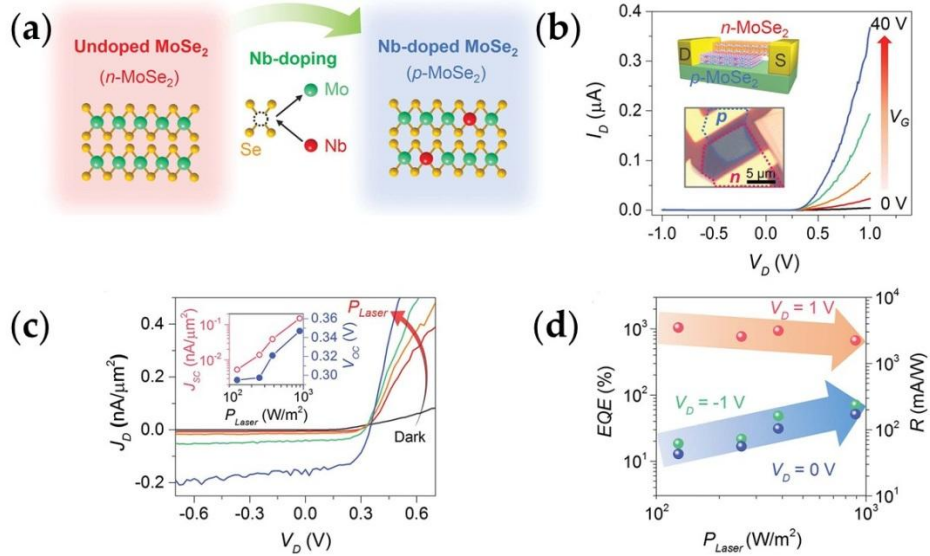

**Fig. 3.** (a) Schematic diagram of Nb-doping for converting n-$MoSe_2$ to p-$MoSe_2$. (b) $I_{ds}$ - $V_{ds}$ characteristics measured in different gate configurations. The insets: optical image and schematic of the vertical homojunctions based on $MoSe_2$. (c) $J_d$- $V_{ds}$ curves under different laser powers. The inset: short circuit current and open circuit voltage as a function of $P_{Laser}$. (d) EQE and D* with respect to the laser power under different $V_{ds}$.[50]

The fabricated p-n junction exhibits typical rectification properties, as shown in

Fig. 3b. Not only that, the ideality factor was calculated to be ~1, which implies the device is an ideal diode. With increasing illuminated light intensity, short-circuit current ($I_{sc}$), open-circuit voltage ($V_{oc}$) and reverse current are all increased, as shown in Fig. 3c. This is consistent with the photovoltaic mechanism where the generated photocurrent can add to the reverse current. In Fig. 3d, both the EQE and the responsivity are enhanced at $V_{ds} = 0$ and $V_{ds} = -1$ V as the laser power increased. Photocarriers can be more easily drifted to electrode across the depletion region under reverse bias. For the vertical MoSe$_2$ homojunction by elemental doping, the maximum EQE and R value can reach ~1049% and be more than 1 A·W$^{-1}$ at $V_{ds} = 1$ V, $P_{Laser} = 890$ W·m$^{-2}$, respectively.

The elemental doping is superior in terms of versatility and stability over surface chemical treatment based on absorption or intercalation of volatile species. However the elemental doping is more suitable for TMDC bulk materials rather than very few layered materials. In order to fabricate PN junctions based on the few layer TMDCs, mechanical exfoliation is often essential, which can induce uncertainty factors such as the thicknesses and the resistance between the flakes. Thus the current method is not suitable for practical application. Recently Kim *et al*. realized phosphorus doping in monolayer and bilayer MoS$_2$ by laser.[54] This approach can be controlled effectively by varying the laser irradiation power and time. But more solutions of elemental doping in thin layers are still needed. We expect to realize PN homojunctions based on layered TMDCs through the elemental doping during the CVD growth, which is important for the application of TMDCs.

### 3.3. Homojunction fabricated by electrostatic gating

Unlike other TMDCs such as MoS$_2$, MoSe$_2$ and WS$_2$, WSe$_2$ has been demonstrated to be ambipolar transport.[40-42,55] The carrier type of WSe$_2$ can be controlled through electrostatic gating. The homojunction based on WSe$_2$ has been created successfully by this method.[40]

The schematic diagram of the locally-gated junction is presented in Fig. 4a. The single-layer WSe$_2$ flake was transferred on two local gates covered with BN as a dielectric layer. Except BN,[40,42,56] conventional dielectrics such as HfO$_2$,[41] Si$_3$N$_4$,[43] Al$_2$O$_3$[39] have also been used in similar type devices. Compared with HfO$_2$ and Si$_3$N$_4$, BN flake has atomically flat surface and charge traps-free interface, it can improve the interface between BN and WSe$_2$, thus enhance the mobility of devices.

In Fig. 4b, the ambipolar $I_{ds}$ can be controlled by $V_g$, implying both hole and electron-doping can be accessed. This phenomenon can be attributed to that the Femi level $E_F$ can be effectively moved up and down across the bandgap by $V_g$ in thin layers of WSe$_2$. The moving of $E_F$ to the conductive band or valance band will induce different carrier types and concentrations. Similar behavior has also been reported by others.[42,43] So the types of the junctions (PP, PN and NN) can be adjusted by controlling $V_{lg}$ and $V_{rg}$, as shown in Fig. 4c.

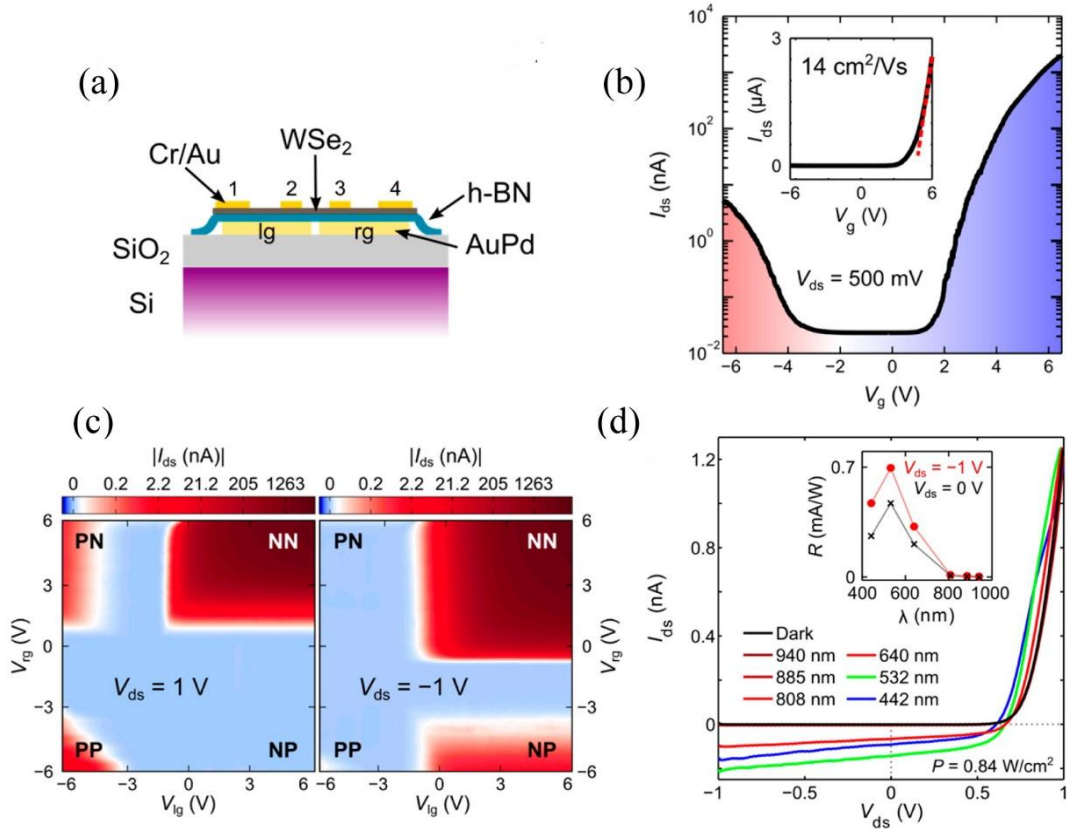

**Fig. 4.** (a) Device schematic diagram. (b) The log plotted $I_{ds}$-$V_g$ curve was measured under $V_{ds} = 500$ mV. The color gradient implies the transition between hole- and electron-doping. Inset shows the data on a linear scale. (c) Color maps of $I_{ds}$ under different $V_{lg}$, $V_{rg}$ and $V_{ds}$ which suggest the types of the junctions. (d) $I_{ds}$-$V_{ds}$ characteristics under different illumination wavelength. The inset shows the responsivity (R) under $V_{ds} = -1$ V and $V_{ds} = 0$.[40]

Then we focus on the photoresponse of the PN junction formed by electrical gating. The maximum responsivity is 0.7 mA·W$^{-1}$ at $V_{ds} = -1$ V and 532 nm illuminations. Meanwhile the EQE reach a maximum value of 0.1%. In addition, the response time is in the order of 10 ms. Though the EQE here is lower compared with the two doping methods mentioned above, locally electrostatic gating is still a novel way to tune the junctions and control their electrical properties.

To conclude, the homojunction photodetectors based on TMDCs have been introduced, which can be realized by three different ways: surface chemical doping, elemental doping and electrostatic gating. Compared with intrinsic TMDCs, photodetectors based on homojunctions can largely improve the EQE and R. Meanwhile homojunctions are fabricated by only one kind of TMDCs, which makes them promising as detectors for special wavelength applications from visible to near infrared.

## 4. Photodetectors based on heterojunctions of TMDCs

In contrast to homojunctions, heterojunction is the semiconductor interface that occurs between two dissimilar materials with unequal bandgaps. Flakes of TMDCs can be placed on conventional semiconductor materials, organic semiconductor or on other 2D materials by ways such as CVD, mechanical exfoliation and transfer and so on. Due to the unique layer thickness dependence of the bandgaps, the heterojunctions can also be made of only single TMDCs with two different layer thickness.

### 4.1. Heterojunctions based on TMDCs, conventional bulk semiconductors and organic semiconductors

**Heterojunctions based on TMDCs and conventional bulk semiconductors.** Compared with heterojunctions based on two different 2D materials, it is easier and more feasible to combine a 2D layered material with conventional semiconductor materials such as $MoS_2/Si$,[57-62] $WSe_2/InAs$,[63] $MoS_2/GaAs$[64] and $MoS_2/GaN$.[65]

The $MoS_2/Si$ heterojunctions have been widely investigated.[57-62] Most of such junctions were fabricated by mechanical exfoliation and transfer.[57-60] Esmaeili-Rad *et al*. fabricated heterojunction of $MoS_2/a$-Si and they obtained the photoresponsivity of 210 mA/W at green light.[57] Vertical $MoS_2/p$-Si heterojunction was fabricated by transferring and the high photoresponsivity of 7.2 A/W was obtained.[60] In order to have large scale such heterojunctions, Wang *et al*.[61] and Hao *et al*.[62] fabricated the $MoS_2/Si$ heterojunctions by magnetron sputtering.

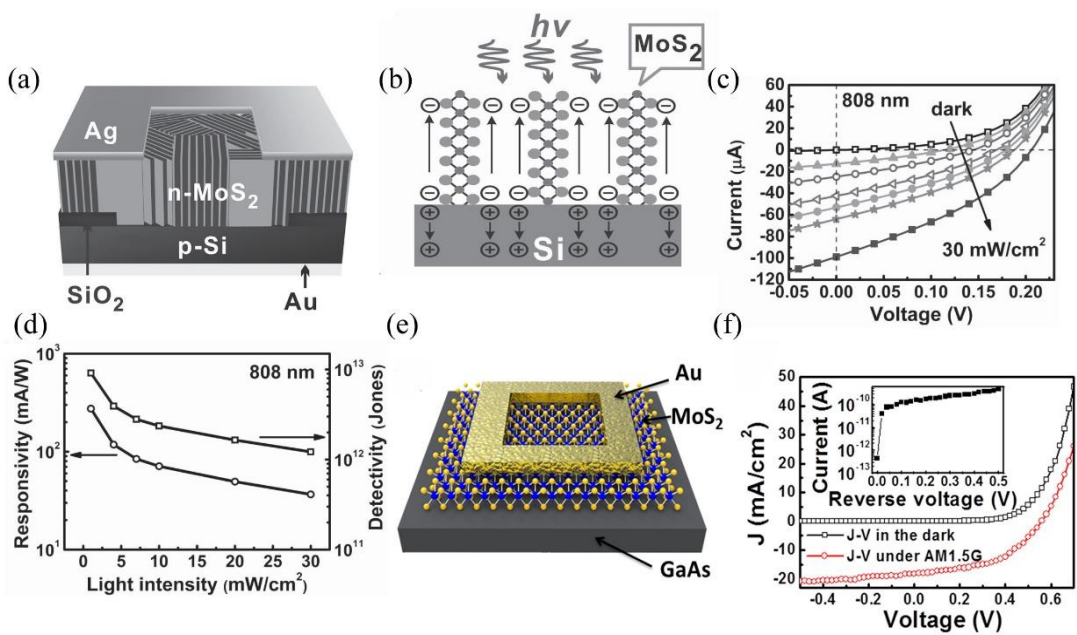

**Fig. 5.** (a) Device schematic of the $MoS_2/Si$ heterojunction photodetector.[61] (b) Schematic illustration shows the separation of the photocarriers under illumination.[61] (c) Photoresponse characteristics measured under different light intensities of 808 nm laser.[59] (d) Responsibility and specific detectivity with respect to the light intensity.[61] (e) Schematic structure of $MoS_2/GaAs$ heterojunction.[64] (f) Current density-voltage curves of $MoS_2/GaAs$ heterojunction under dark and illumination.[64]

Fig. 5a shows the device geometry fabricated by Wang *et al*.[61] The $MoS_2$ film was deposited on the p-Si substrate with a predefined $SiO_2$ window by magnetron sputtering. The deposited molecular layers of $MoS_2$[61] are perpendicular to the substrate as shown in Fig. 5b, which is different from the $MoS_2$ prepared by mechanical exfoliation[57,59,60] and chemical vapour deposition.[58]

The $I_{sc}$ and $V_{oc}$ increase with the increased light intensity in Fig. 5c, which strongly suggests that photocarriers are separated by the electric field at the interface of $MoS_2$ and Si. Responsivity can reach ~300mA·W$^{-1}$ under the light intensity of $1mW·cm^{-2}$ which is comparable to most of the devices based on $MoS_2$[26,50,57,66]. The maximum value of specific detectivity is ~$10^{13}$ Jones at 1 mW·cm$^{-2}$ which is three orders of magnitude higher than $MoS_2$/Si heterojunction photodetectors by transferring.[60]

The rising and decaying time of the $MoS_2$/Si heterojunctions were calculated to be 3 μs and 40 μs, which is by far the reported fastest response in TMDCs heterojunctions. The fast response is directly related with the structure and can be attributed to the following reasons: the strong build-in electric field which can efficiently separate photocarriers; the vertically standing layered structure of $MoS_2$ layers which offers high-speed paths; thin $SiO_2$ between Si and $MoS_2$ which serves as a passivation layer to reduce the interface defects.

Lately Xu *at el.* developed $MoS_2$/GaAs heterojunctions which have wide response band width from ultraviolet to visible light.[64] In order to decrease the dark current, h-BN was inserted into the interface between $MoS_2$ and GaAs. Owing to h-BN, the detectivity can reach $1.9 \times 10^{14}$ Jones which is even higher than that of $MoS_2$/Si heterojunction.[61] $MoS_2$/GaN heterojunctions have also been prepared by Ruzmetov *at el.* via power vaporization technique and show very high-quality van der Waals interface.[65]

The $MoS_2$/Si heterojunctions show not only high specific detectivity but also short response time.[61] The photodetectors based on $MoS_2$/GaAs heterojunctions have wider photoresponse band width. The efforts above confirm that combining 2D TMDCs with conventional semiconductor technology is a promising way to make high performances PN heterostructural photodetectors.

**Heterojunctions based on TMDCs and organic semiconductor.** Compared with conventional bulk semiconductor, much cheaper organic semiconductor materials including pentacene,[67,68] $C_8$-BTBT,[69-71] CuPc[72] and PTCDA[73] have also enriched the design of heterojunction based on TMDCs.

A lot of efforts have been made to epitaxially grow organic semiconductor thin film on 2D materials such as graphene, BN and $MoS_2$. He *et al.* grew high-quality few layer $C_8$-BTBT on graphene and BN via van der Waals epitaxy. The thickness of $C_8$-BTBT can be well controlled down to monolayer.

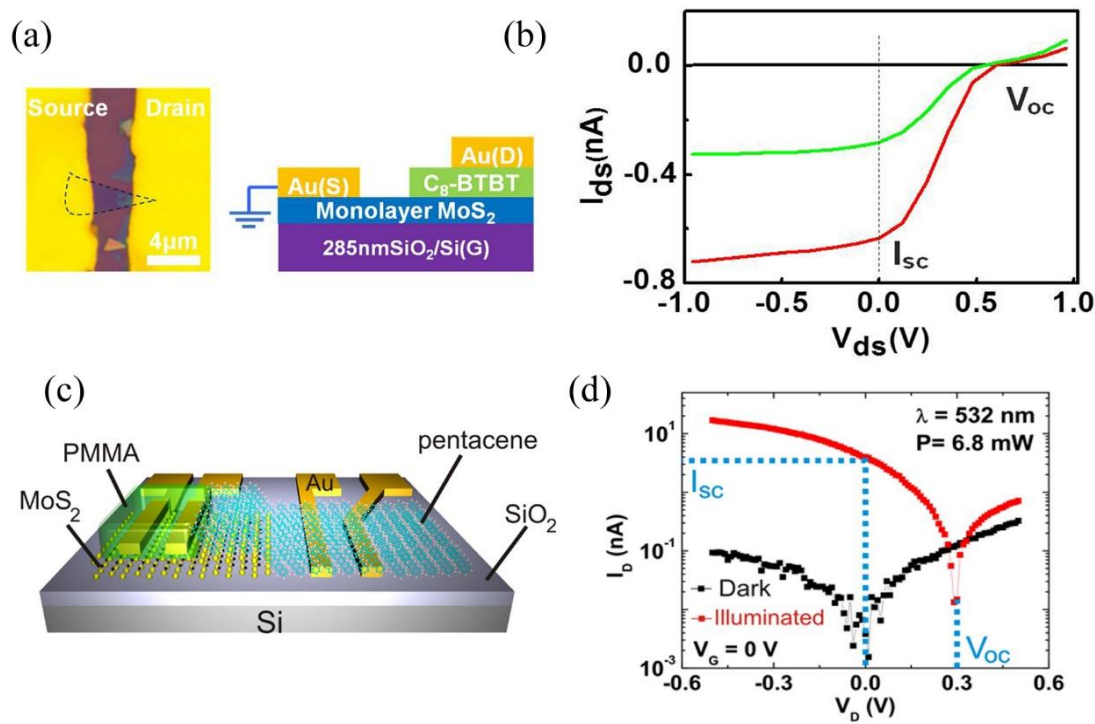

**Fig. 6.** (a) Microscope image and device schematic of vertical $C_8$-BTBT/ $MoS_2$ heterojunction.[70] (b) $I_{ds}$-$V_{ds}$ under white light illumination with different power.[70] (c) Structure of pentacene/ $MoS_2$ heterojunction.[67] (d) $I_{ds}$-$V_{ds}$ under dark and illumination.[67]

Lately the heterojunctions based on monolayer $MoS_2$ and p-type $C_8$-BTBT in Fig. 6a were also presented by He *et al.*[70] $MoS_2$ was mechanically exfoliated on $SiO_2$/Si as an epitaxy substrate and then few-layer $C_8$-BTBT on $MoS_2$ was grown by epitaxal growth. The sharp and clean interface can be formed between the $MoS_2$ and the organic $C_8$-BTBT, which is important for the performance of the heterojunction. Under illumination, obvious $I_{sc}$ and $V_{oc}$ response are observed in the heterojunction as shown in Fig. 6b and the responsibility can reach 22 mA/W under photovoltaic mode. At the same time, Jariwala *et al.* fabricated heterojunctions based on $MoS_2$ and p-type pentacene.[67] As shown in Fig. 6d, the devices show apparent diode-like behavior and large photoresponse. This kind of hybrid heterojunctions has attracted a lot of attention and can show good performances by selecting appropriate device structures and materials. Compared with conventional semiconductors, organic semiconductors can be made in inexpensive methods. Moreover the low annealing temperature of organics allows for flexibility of substrate choice. It is promising to fabricate cheaper photodetectors by combining TMDCs and organic semiconductors.

### 4.2. Photodetectors based on heterojunctions of only TMDCs

**Fabrication of heterojunctions based on only TMDCs.** Because of the weak van der Waals interactions, we can stack any two kinds of 2D materials together in order to form heterojunctions by mechanical exfoliation and transfer regardless of the lattice mismatch, which will gives us wide variety selections of layered materials. A lot of efforts have been made to stack vertical heterojunctions of 2D TMDCs by mechanical exfoliation and transfer as shown in heterojunctions such as $MoS_2/WS_2$[74,75], $MoS_2/WSe_2$,[76-82] $MoS_2/MoSe_2$,[83] $MoS_2/MoTe_2$[84] and $MoSe_2/WSe_2$[85] and so on. Although such vertical heterojunctions show very good performance in photodetection, the yield and efficiency of making such devices is very low.

In order to reduce the steps of mechanical exfoliation and transfer, Cheng *et al.* combine physical vapor deposition process, mechanical exfoliation and transfer together.[86] But the efficiency and the quality of junctions still have not been improved much.

Since $MoS_2$ layers were grown by chemical vapor deposition (CVD),[87-89] CVD has been regarded as a feasible and efficient way to directly fabricate heterojunctions on a large scale. Moreover CVD-grown devices showed stronger interlayer interaction compared with their transferred counterparts.

In 2014, Gong *et al.*[90] grew both vertical and lateral $MoS_2/WS_2$ heterojunctions by CVD. As the FET, the vertical $MoS_2/WS_2$ heterojunctions have very high ON/OFF ratio of $10^6$ and the estimated mobility is in the range of 15 to 34  $cm^2 \cdot V^{-1} \cdot s^{-1}$, which is faster than $MoS_2/WS_2$ made by mechanical transfer (0.51 $cm^2 \cdot V^{-1} \cdot s^{-1}$ ). The high mobility can be attributed to the clean interface between $WS_2$ and $MoS_2$, which reduces the unwanted defects between layers to improve the charge transfer. Meanwhile the lateral $MoS_2/WS_2$ heterojunctions have been demonstrated to be intrinsic monolayer PN junctions without external electrostatic gating.

Afterwards scalable production of few-layer $MoS_2/WS_2$ vertical heterojunction array was reported by Xue *et al.*[91]

The few-layer $MoS_2/WS_2$ vertical heterojunctions (Fig. 7a) were fabricated by sulfurization of patterned $WO_3$ and Mo sheets with an ambient pressure thermal reduction process. There are two Raman-active modes ($^1E_{2g}$, $A_{1g}$) that can be generally observed due to the selection rules for scattering radiation or limited rejection of the Rayleigh scattering radiation.[93] In the Raman spectra of $MoS_2/WS_2$, the position of $^1E_{2g}$ and $A_{1g}$ peaks do not change, implying $MoS_2$ and $WS_2$ to be individual sheets. So the two-step chemical vapor deposition approach has been demonstrated to effectively prevent the mixture of $MoS_2$ and $WS_2$, resulting in a well-defined interface between $MoS_2$ and $WS_2$. Moreover, the thicknesses of $MoS_2$ and $WS_2$ in the heterojunctions can be adjusted by controlling the thicknesses of the predeposited $WO_3$ and Mo sheets during the fabrication.

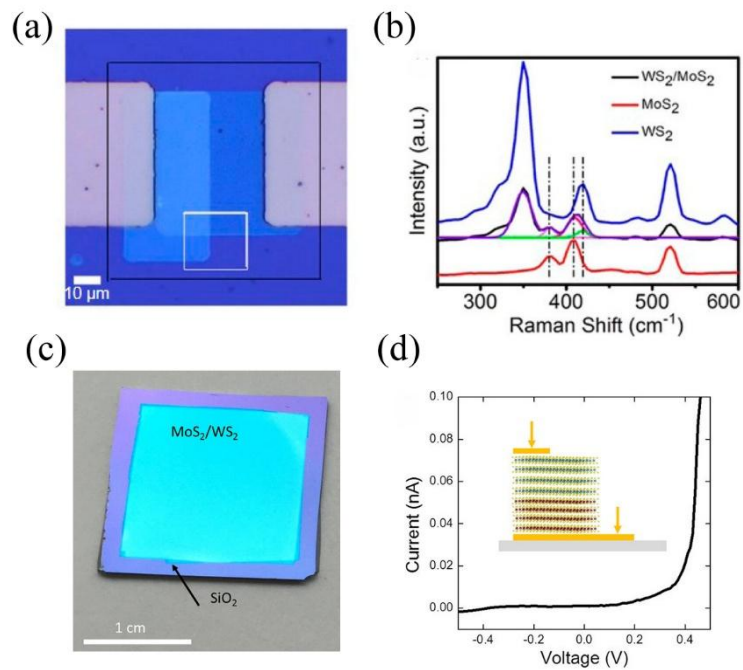

**Fig. 7.** (a) Optical micrograph of the MoS$_2$/WS$_2$ vertical heterojunction device.[91] (b) Raman spectra of MoS$_2$, WS$_2$ and MoS$_2$/WS$_2$.[91] (c) Optical image of vertical MoS$_2$/WS$_2$ heterojunctions grown by chemical vapor deposition.[92] (d) I-V measurement across a MoS$_2$/WS$_2$ heterointerface.[92]

In the same year, Li *et al.* realized the controlled epitaxial growth of lateral WSe$_2$/MoS$_2$ heterojunction.[94] WSe$_2$ was grown on c-plane sapphire substrates through van der Waals epitaxy and then MoS$_2$ was epitaxially grown at the edge of WSe$_2$. The atomically sharp transition in compositions at junctions can be achieved through precisely controlling the two-step epitaxial growth. Very recently, Choudhary *et al.*[92] reported large-area (>2 cm$^2$) vertical MoS$_2$/WS$_2$ heterojunctions grown by chemical vapor deposition. The heterojunctions show diode-like current rectification, which means them to be high-quality van der Waals PN junctions (Fig. 7d).

To conclude, although mechanical exfoliation and transfer are still widely used to stack two kinds of 2D materials, CVD techniques have been demonstrated to be a reliable way to fabricate large scale vertical and lateral heterojunctions of TMDCs which are more promising for application. Meanwhile some new methods also appear. Tan *et al.* achieved 2D semiconductor hetero-nanostructures by the epitaxial growth in liquid phase at room temperature.[95] Mahjouri-Samani *et al.* fabricated array of lateral MoSe$_2$/MoS$_2$ heterojunction by lithographic patterning and controllable conversion of exposed MoSe$_2$ to MoS$_2$, which also provides some new ideas for device fabrication.[96]

**Photoelectric properties of vertical heterojunctions based on only TMDCs.** Furchi *et al.*[77] fabricated vertical stacks of WSe$_2$/MoS$_2$ exactly by mechanical exfoliation and transfer. The EQE is a low value of about 1.5%. A much higher performance in similar vertical WSe$_2$/MoS$_2$ heterojunctions was reported by Cheng *et al.*(Fig. 8a).[86]

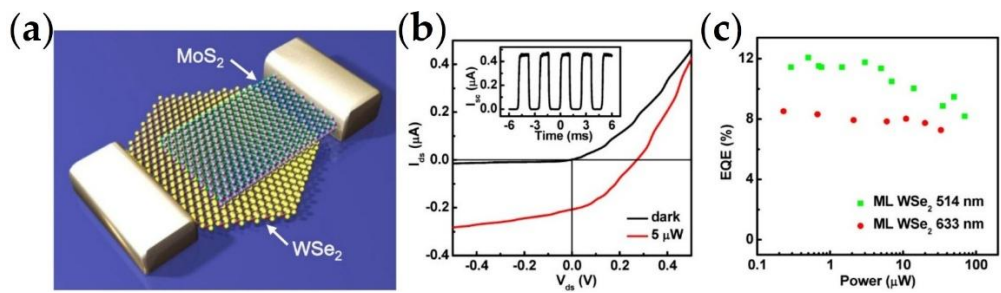

**Fig. 8.** (a) Schematic of vertical WSe$_2$/MoS$_2$ heterojunction; (b) I$_{ds}$-V$_{ds}$ curves measured in the dark and under the illumination whose wavelength is 514 nm and power is 5 μW. The inset shows response of the current with respect to periodic on/off of the illumination; (c) EQE under different power of 514 nm and 633 nm laser at V$_{ds}$ = 0 and V$_g$ = 0.[86]

The vertical WSe$_2$/MoS$_2$ heterojunctions have obvious open-circuit voltage and short-circuit current under laser illumination (514 nm, 5 μW) (Fig. 8b). Also a fast response time of 100 μs has been demonstrated. The maximum value of EQE can reach 12% under laser illumination (514 nm, 0.5 μW) (Fig. 8c) which is higher than homojunctions based on WSe$_2$.[41,43] The high performance can be resulted from the efficient separation of photocarriers at the interface of WSe$_2$ and MoS$_2$.

More recently, in the MoS$_2$/WSe$_2$ heterojunction, the ultrafast electron transfer from WSe$_2$ to MoS$_2$ sheet was demonstrated to be 470 fs and a 99% efficiency upon optical excitation. This implies MoS$_2$/WSe$_2$ heterojunction could be have big potential for high-speed photodetectors.[82]

**Photoelectric properties of lateral heterojunctions based on only TMDCs.** In addition to vertical heterojunctions, lateral 2D heterojunctions have also emerged. Lateral heterojunctions based on TMDCs are commonly fabricated such as MoSe$_2$/MoS$_2$,[96,97] MoSe$_2$/WSe$_2$,[98] WS$_2$/MoS$_2$,[90] WS$_2$/WSe$_2$,[97] WSe$_2$/MoS$_2$.[94]

Duan *et al.* realized the lateral epitaxial growth of WS$_2$/WSe$_2$ heterojunctions (Fig. 9a).[97] Afterwards, a 50 nm-thick Al$_2$O$_3$ was deposited on part of WSe$_2$ to insulate it from the contact for WS$_2$. Compared with the I$_{ds}$-V$_{ds}$ curve in the dark, the heterojunctions exhibited obvious photovoltaic with an open-circuit voltage of 0.47 V and short-circuit current of 1.2 nA (Fig. 9b). Moreover, the response time is less than 100 μs which is two orders of magnitudes less than the response time of the lateral WSe$_2$ homojunction.[40] The faster response speed is originated from the carriers transport mainly in surface of lateral heterojunctions. The value of EQE is about 9.9% for these devices. Li *et al.* investigated the WSe$_2$/MoS$_2$ lateral heterojunctions.[94] They found that the depletion width of the heterojunction is 320 nm and device shows clear photovoltaic effect.[94]

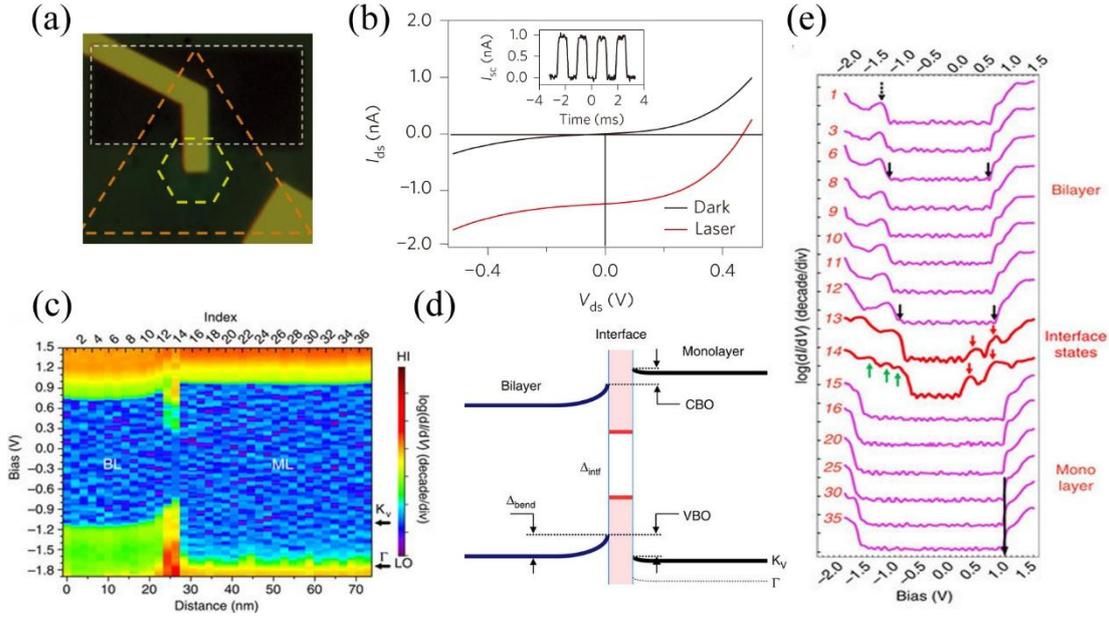

**Fig. 9.** (a) Optical image of lateral WS$_2$/WSe$_2$ heterojunction.[97] (b) I$_{ds}$-V$_{ds}$ characteristics in the dark and under laser illumination (514 nm 30 nW). The inset shows the photocurrent response under periodic on/off laser illumination.[97] (c) Colour-coded rendering of the real space imaging of band profile plotted in terms of log(dI/dV) in monolayer-bilayer WSe$_2$.[99] (d) Band alignment in monolayer-bilayer WSe$_2$ heterojunctions.[99] (e) Selective subset of log(dI/dV) spectra. The green and red arrows represent valance and conduction band of the interface states.[99]

Moreover, due to the thickness dependence of bandgap and band edges of TMDCs, a novel heterojunctions have appeared which consist of different layers of the same TMDCs. Interlayer states can exist in these heterojunctions such as monolayer-bilayer WSe$_2$ heterojunctions,[99] monolayer-multilayer MoS$_2$ heterojunction.[100,126] Zhang *et al.* studied interlayer states between monolayer and bilayer WSe$_2$ by low-temperature scanning tunneling micrology and spectroscopy (STM/S). The interlayer states have a gap value of 0.8eV as shown in spectrum #13 and spectrum #14 of Fig. 9e. Because of the interlayer states, the band bending and build-in electric field can be induced between interfaces (Fig. 9d), which implies them to be a novel class of heterojunctions for photodetectors. The advantage of this kind of heterojunctions is that gaps can be adjusted by thickness. Thus through the combination of same TMDCs of different thickness, detecting wavelength can be tuned and extended.

To conclude, along with development of different techniques, large scale vertical and lateral heterojunctions have been demonstrated and they have exhibited some excellent properties such as the fast response time which is less than 100 μs. Heterojunctions which consist of different layers of the same TMDCs are also greatly promising for new design of photodetectors. The different structures provide a more solid foundation for more complex structures such as PIN junctions and quantum well structures. But it is still not good enough for applications. For example the area of heterojunctions with higher quality is not larger enough and EQE is still very low. So the fabrication and performances of TMDC/TMDC heterojunctions all need to be improved.

### 4.3. Photodetectors based on heterojunctions of TMDCs and other 2D materials

Though TMDCs have the advantage of strong light absorption, applications based on TMDCs are limited by the natures such as the low carrier mobility compared with graphene[2,5,101] and few-layer black phosphorus.[102-104] Thus in order to make devices with better performance, a native way is to combine good properties of TMDCs and the advantages of other 2D materials by stacking their layers together.

The approach has already been demonstrated by Deng *et al*.[105] The $MoS_2$/BP heterojunctions combine high carrier mobility of BP with strong light absorption of $MoS_2$. The heterojunctions show high responsivity of 418 mA·W$^{-1}$, but low EQE of 0.3%. Due to built-in electric field, the photocarriers can be effectively separated and then photoelectrons are injected into BP layer, rather than trapped in $MoS_2$.

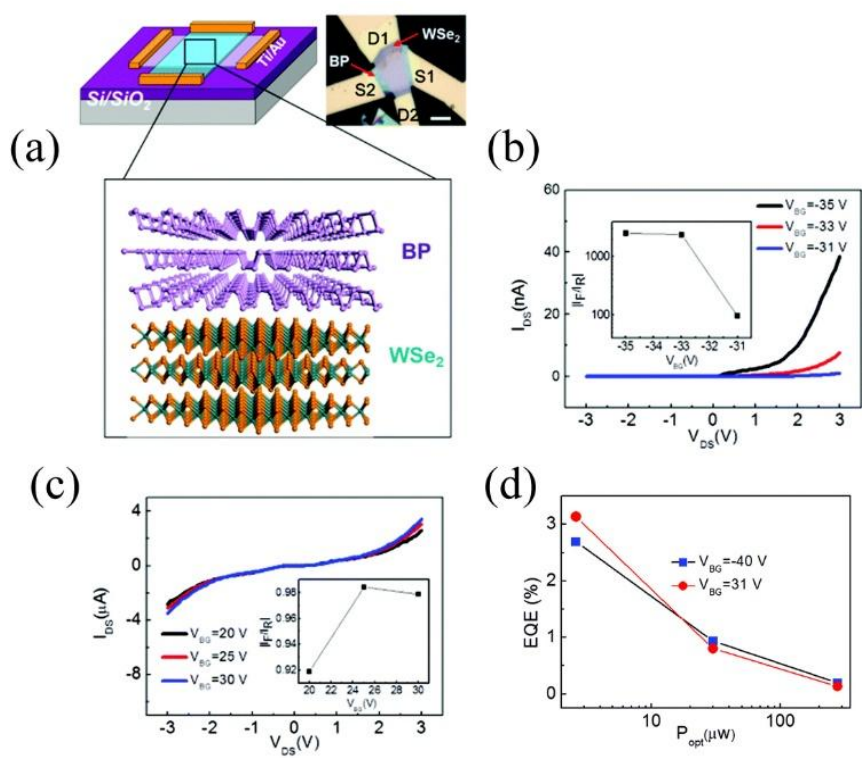

**Fig. 10.** (a) Device schematic diagram and optical image of $WSe_2$/BP heterojunction. (b) and (c) $I_{ds}$ −$V_{ds}$ characteristics under different $V_g$. The insets show the rectification ratio with respect to $V_g$. (d) shows the relations between EQE and power of laser under $V_g$ = -40 V and $V_g$ = 31 V.[106]

More recently, the similar study of $WSe_2$/BP heterojunctions (Fig. 10a) were reported by Chen *et al*.[106] Due to the weak screening effect and the ambipolar transport of $WSe_2$[40-42,55] and BP,[103,107] different types of heterojunctions can be well controlled by electrostatic gating. As shown in Fig. 10b and 10c, the rectification properties have almost disappeared when $V_{ds}$ > -33 V and $V_{ds}$ < 25 V. The values of EQE have been calculated in Figure 8d and the maximum value is 3.1% at $V_g$ = 31 V, which is larger than 0.3% in the $MoS_2$/BP heterojunctions.[105] By stacking TMDCs and other 2D materials together, their merits can be combined in the same device such as the bipolar transport of WSe$_2$ and BP, light absorption of MoS$_2$ and high mobility of graphene and so on.[108-111]

Although the junctions have shown excellent properties, there are still many methods to improve performances by using graphene as electrodes to reduce the influence of barrier,[112,113] facilitating the carrier collection by designing a vertical sandwiched structure[76,113,114] and using BN as a dielectric layer to improve interface and increase the reflection between the layers.[56,115]

Early Luo *et al* has proven the response time of photodetectors based on InSe can be decreased to 120 μs by using graphene as electrodes, which is about 40 times faster than the detectors using metal electrodes.[112] Lee *et al*. fabricated vertical MoS$_2$/WSe$_2$ heterojunction by using top and bottom graphene electrodes (Fig. 11a).[76] The device had photoresponsivity of ~10 mA·$W^{-2}$, which is five times larger than that of MoS$_2$/WSe$_2$ heterojunction with metal contacts. There are several advantages to use graphene as electrodes. Firstly ultrafast carrier mobility of graphene is very useful in the separation of photogenerated carriers, which can enhance the photoresponsivity and further improve detectivity. Secondly the graphene can absorb few light which has little effects on the light absorption of the heterostructure. Finally the overlapping of the top graphene and bottom graphene facilitates the carrier collection by reducing the vertical transport channel, which can also improve the response speed. So using graphene as electrodes to improve the performances of photodetectors based on TMDCs and other 2D materials was demonstrated to be an effective way.

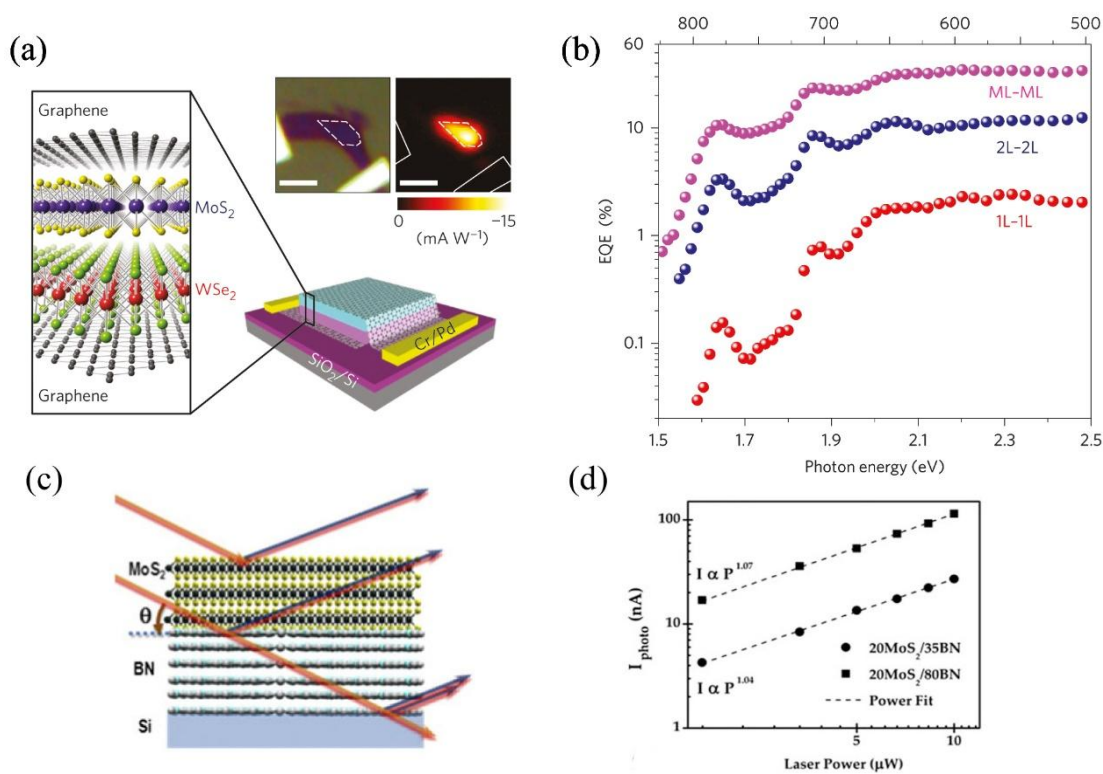

**Fig. 11.** (a) The schematic diagram of MoS$_2$/WSe$_2$ junction sandwiched between top and bottom graphene electrodes.[76] (b) shows the relationship between EQE and excitation energy.[76] (c) the possible existence of the interfaces between materials. (d) the photocurrent with respect to laser power.[115]

Except using graphene as electrodes, in order to improve the performances of TMDCs, Wasala *et al.* investigated the effects of BN on the optical properties of $MoS_2$.[115] As shown in Fig. 11d, the photocurrent increased with increasing the thickness of the underlying BN layer. Because the BN layers can increase the reflections between layers, which can result in the enhancement of photocurrent. It implies BN also can be used to enhance the photocurrent response properties of TMDCs such as photoresponsivity and detectivity. Moreover because the h-BN has atomically flat and smooth surface with lack of dangling bonds, it has been widely used as an ideal subatrate and dielectric layer for 2D materials such as graphene and TMDCs.[56,116,117]. The h-BN can also be used as intercalation which could decrease the dark current and further improve detectivity.[61]

In conclusion, completely vertical structure can reduce the transport channel length, which can greatly enhance the photoresponse speed of photodetectors. Moreover, using graphene electrodes and BN reflection layer have been demonstrated to be promising methods to improve photocurrent response. Combing those methods, high-speed and highly sensitive photodetector could be realized by stacking of TMDCs and other two dimensional materials.

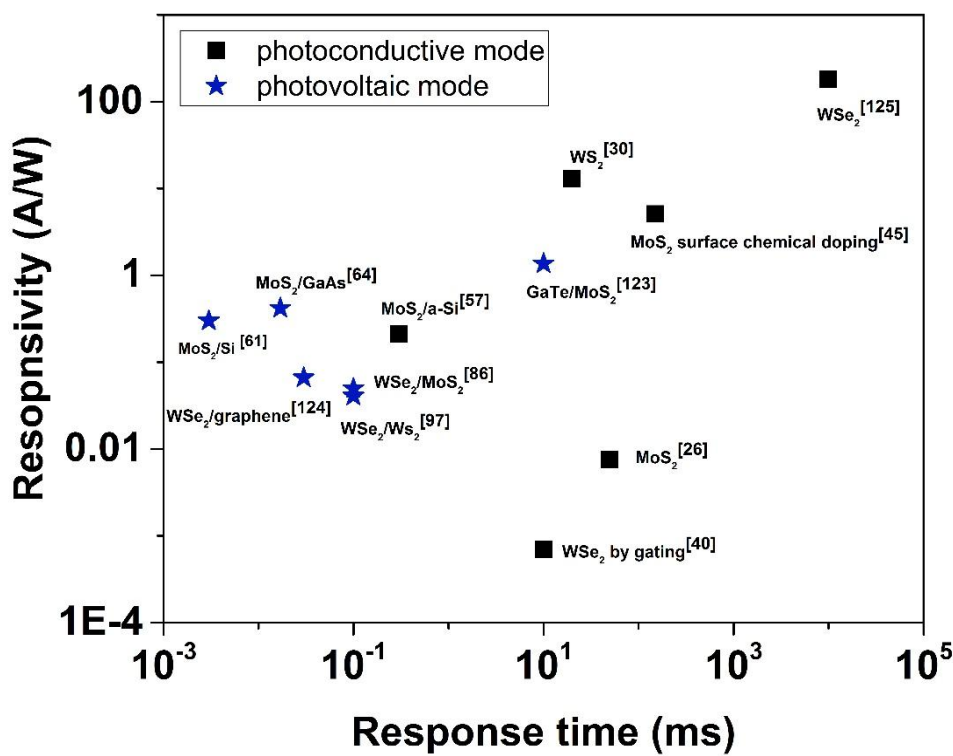

**Fig. 12.** Responsivity against response time for some devices reviewed.

Fig. 12 and table 1 summarized some experimental results from 2D material photodetectors in the review. The responsivity and response time for part of the reviewed devices are summarized in Fig. 12. In table 1, we list the configuration, the fabrication technology, EQE, responsivity, specific detectivity and rise time. All those present the performances of the detectors.

**Table 1.** Summary of literature data for photodetectors based on two-dimensional TMDCs. The table contains the configuration, the fabrication technology, mode and some benchmark parameters (external quantum efficiency (EQE), responsivity (R), specific detectivity (D*), rise time ($\tau_r$))

| Material (Configuration) | Technology | Mode | Figures of merit | | | | Ref. |
|---|---|---|---|---|---|---|---|
| | | | EQE (%) | R (A/W) | D* (Jones) | $\tau_r$ (ms) | |
| Single-layer $MoS_2$ (FET) | Mechanical exfoliation | photoconductive | NA | 0.0075 | NA | 50 | [26] |
| | | photoconductive | NA | 880 | NA | >$10^4$ | [31] |
| Multilayer $MoS_2$ (FET) | Mechanical exfoliation | photoconductive | NA | ~0.12 | ~$10^{11}$ | NA | [53] |
| $MoS_2$ (Vertical Homojunction) | Plasma doping | photovoltaic | 78 | NA | NA | NA | [52] |
| $MoS_2$ (Lateral homojunction) | Surface chemical doping | photoconductive | 1200 | 5.07 | $3 \times 10^{10}$ | 100-200 | [46] |
| | | photoconductive | 81 | 0.33 | $1.6 \times 10^{10}$ | NA | |
| n-$MoSe_2$/ p-$MoSe_2$ (Vertical homojunction) | Elemental doping, mechanical exfoliation and transfer | photoconductive | 1049 | ~3 | NA | NA | [50] |
| $WSe_2$ (Lateral homojunction) | Electrostatic gating, mechanical exfoliation and transfer | photoconductive | 0.1 | $7 \times 10^{-4}$ | NA | ~10 | [40] |
| | | photovoltaic | 0.2 | NA | NA | NA | [41] |
| $MoS_2$/Si (Heterojunction) | Magnetron sputter | photovoltaic | NA | 0.3 | $10^{13}$ | $3 \times 10^{-3}$ | [61] |
| | Mechanical exfoliation and transfer | photoconductive | NA | 7.2 | ~$7 \times 10^9$ | NA | [60] |
| $MoS_2$/GaAs (Heterojunction) | CVD | photovoltaic | NA | 0.419 | $1.9 \times 10^{14}$ | 0.017 | [64] |
| $C_8$-BTBT/ $MoS_2$ (Heterojunction) | Mechanical exfoliation and epitaxial growth | photovoltaic | NA | 0.022 | NA | NA | [70] |
| $WSe_2$/ $MoS_2$ (Vertical heterojunction) | PVD and mechanical exfoliation and transfer | photovoltaic | 12 | NA | NA | < 0.1 | [86] |
| $WS_2$/ $WSe_2$ (Lateral heterojunction) | CVD | photovoltaic | 9.9 | NA | NA | < 0.1 | [97] |
| $MoS_2$/BP (Vertical heterojunction) | CVD and mechanical exfoliation and transfer | photovoltaic | 0.3 | NA | NA | NA | [105] |
| | | photoconductive | NA | 0.418 | NA | NA | |
| $WSe_2$/BP | Mechanical exfoliation | | | | | | |

| ( Vertical heterojunction) | and transfer | photovoltaic | 3.1 | NA | NA | NA | [106] |

## 5. Conclusion and perspectives

This review provides an overview of the recent development in photodetectors based on junctions of TMDCs from structures, fabrications and performances. Owing to the built-in electric field in those junctions, the photogenerated electrons and holes can be more effectively separated. So compared with photodetectors based on FET type of intrinsic TMDCs, photodetectors based on junctions of TMDCs show many advantages (such as lager responsivity or faster response speed) as shown in Table 1 and Fig. 12. Moreover, selection of the materials, the device fabrication processing all have great influences on performances of photodetectors.

Homojunctions can be created by the following three methods: surface chemical doping, elemental doping and electrostatic gating. Although surface chemical doping is suitable for thin layers of TMDCs, the homojunctions based on the surface chemical doping are easy to be affected by the external environment. Elemental doping is a fesiable way for vertical stable homojunctions with high performance. Although electrostatic gating can also be used to create homojunctions, the EQE of these junctions is still very low. Because of the limination of charge mobility, lateral homojunctions fabricated by surface chemical doping and electrostaic show slow photoresponse. However homojunctions are based on the similiar materials, which makes them easier to realize the sensitive detection of special wavelength.

Heterojunctions can be fabricated by mechanical exfoliation and transfer, CVD and some other methods. Compared with mechanical exfoliation and transfer, CVD method has been demonstrated to be the most effective and promising method to obtain large scale heterojunctions based on TMDCs with atomically sharp interface. Moreover CVD-grown junctions showed stronger interlayer interaction compared with transferred counterparts.

In vertical heterojunctions, using graphene as top and bottom electrodes can effectively shorten the vertical transport distance of photogenerated carriers, which reduces the response time by several orders of magnitude. The photoresponse can be stongly enhanced using BN as the reflection layer. Taking the advantages of graphene and BN, the heterojunciton photodetectors should have potential appliacations in high speed and weak signal photodetections.

Photodetectors with wide response spectrum are important in many applications. Heterojunctions can combine TMDCs and other semiconductor materials with different absorption wavelength together. $MoS_2$/GaAs heterojunctions have been already demonstrated to have wide spectrum response from ultraviolet to visible light. Thus by selecting two appropriate materials in heterojunctions, it has great potential to realize from ultraviolet to far infrared wider response spectrum photodetectors or multicolor photodetectors.

Moreover, in TMDCs, spin and valley degrees of freedom are intimately coupled due to inversion symmetry breaking together with strong spin-orbit interaction, which has also attracted a great deal of interests.[118] By circularly polarized optical pumping, spin-valley coupled polarization can be generated,[119] which has been observed in optoeletronic devices based on MoS$_2$[120] and WSe$_2$.[121,122] It should be of great significance to introduce spin-polarization and valley-polarization in photodetectors based on junctions of TMDCs.

With the development of TMDCs, different types of junctions based on these materials have been demonstrated in photodetections. But there is still a long way before they can be actually applied. In application, more controllable and effective way to fabricate large scale junctions is highly desired. Also the stability of these junctions still needs to be improved. Moreover, in theory, electronic transport in heterojunctions has been studied for many years by conventional modes, which cannot be applied well in few layer case. A comprehensive understanding of electronic transport in heterojunctions based on 2D materials is urgently needed. Nevertheless these junctions will greatly enrich studies in photodetectors based on the junctions of TMDCs.